\documentclass[aps,prb,twocolumn,superscriptaddress]{revtex4}

\input epsf.sty

\begin{document}


%
%

\title{Effect of a magnetic field on the spin- and charge-density wave order in
La$_{1.45}$Nd$_{0.4}$Sr$_{0.15}$CuO$_{4}$}

\author{S. Wakimoto}
\email[Corresponding author: ]{waki@physics.utoronto.ca}
\affiliation{ Department of Physics, University of Toronto, Toronto,
   Ontario, Canada M5S~1A7 }

\author{R. J. Birgeneau}
\affiliation{ Department of Physics, University of Toronto, Toronto,
   Ontario, Canada M5S~1A7 }

\author{Y. Fujimaki}
\affiliation{Department of Applied Physics, University of Tokyo,
Hongo 7-3-1, Bunkyo, Tokyo 113-8656, JAPAN}

\author{N. Ichikawa\footnote{Present address: Institute for Chemical
Research, Kyoto University, Uji 611-0011, Japan}}
\affiliation{Department of Applied Physics, University of Tokyo,
Hongo 7-3-1, Bunkyo, Tokyo 113-8656, JAPAN}

\author{T. Kasuga}
\affiliation{Department of Applied Physics, University of Tokyo,
Hongo 7-3-1, Bunkyo, Tokyo 113-8656, JAPAN}

\author{Y. J. Kim}
\affiliation{Physics Department, Brookhaven National Laboratory,
   Upton, New York 11973}

\author{K. M. Kojima}
\affiliation{Department of Applied Physics, University of Tokyo,
Hongo 7-3-1, Bunkyo, Tokyo 113-8656, JAPAN}

\author{S.-H. Lee}
\affiliation{ NIST Center for Neutron Research, National Institute of
   Standards and Technology, Gaithersburg, MD 20899 }

\author{H. Niko}
\affiliation{Department of Applied Physics, University of Tokyo,
Hongo 7-3-1, Bunkyo, Tokyo 113-8656, JAPAN}

\author{J. M. Tranquada}
\affiliation{Physics Department, Brookhaven National Laboratory,
   Upton, New York 11973}

\author{S. Uchida}
\affiliation{Department of Applied Physics, University of Tokyo,
Hongo 7-3-1, Bunkyo, Tokyo 113-8656, JAPAN}

\author{M. v. Zimmermann}
\affiliation{Hamburger Synchrotronstrahlungslabor HASYLAB at Deutsches
Elektronen-Synchrotron DESY, Notkestr. 85, 22603 Hamburg, Germany}

\date{\today}

\begin{abstract}

The spin-density wave (SDW) and charge-density wave (CDW) order
in superconducting La$_{1.45}$Nd$_{0.4}$Sr$_{0.15}$CuO$_{4}$ were studied
under an applied magnetic field using neutron and X-ray diffraction
techniques.  In zero field, incommensurate (IC) SDW order appears
below $\sim 40$~K, which is characterized by neutron diffraction
peaks at $(1/2 \pm 0.134, 1/2 \pm 0.134, 0)$.  The intensity of
these IC peaks increases rapidly below $T_{\rm Nd} \sim 8$~K due to
an ordering of the Nd$^{3+}$ spins.
The application of a 1~T magnetic field parallel to the $c$-axis
markedly diminishes the intensity below $T_{\rm Nd}$, while only a slight
decrease in intensity is observed at higher temperatures
for fields up to 7~T.
Our interpretation is that the $c$-axis field suppresses the parasitic
Nd$^{3+}$ spin order at the incommensurate wave vector without disturbing
the stripe order of Cu$^{2+}$ spins. Consistent with this picture, the CDW
order, which appears below 60~K, shows no change for magnetic fields up to
4~T.    These results stand in contrast to the significant field-induced
enhancement of the SDW order observed in superconducting
La$_{2-x}$Sr$_{x}$CuO$_{4}$ with $x\sim0.12$ and stage-4
La$_{2}$CuO$_{4+y}$.  The differences can be understood in terms of the
relative volume fraction exhibiting stripe order in zero field, and the
collective results are consistent with the idea that suppression of
superconductivity by vortices nucleates local patches of stripe order.

\end{abstract}

\pacs{74.72.Dn, 75.30.Fv, 75.50.Ee}

\maketitle

\section{Introduction}

Incommensurate (IC) magnetic correlations are one of the fascinating
characteristics of the hole-doped high-$T_c$ superconducting material
La$_{2-x}$Sr$_{x}$CuO$_{4}$ (LSCO) and related
compounds.~\cite{M.A.Kastner_98}  Dynamic IC correlations in
superconducting LSCO have been observed using neutron scattering
techniques near the optimal doping concentration
$x=0.15$.~\cite{Yoshizawa_88,Bob_89,S.W.Cheong_91}
It was later established that the IC spatial modulation period is
inversely proportional to the optimized superconducting transition
temperature at a given Sr (hole) concentration $x$,~\cite{K.Yamada_98}
suggesting that the incommensurability and superconductivity are
closely related with each other.
On the other hand, {\it static} IC spin correlations have been
extensively studied in La$_{2-x-y}$Nd$_{y}$Sr$_{x}$CuO$_{4}$ (LNSCO).
This was initially because of interest in the so-called 1/8 anomaly,
which refers to the suppression of superconductivity in
La$_{2-x}$Ba$_{x}$CuO$_{4}$ (LBCO) at $x=1/8$, that is accompanied by
the appearance of the low-temperature tetragonal (LTT) $P4_2/ncm$
structure.~\cite{Axe_89}

Nd-doping in LSCO stabilizes the LTT structure
over a wide range of $x$, and significantly suppresses $T_c$.  During
neutron scattering experiments on LNSCO with $0.08 \leq x \leq 0.20$
and $y=0.4$, Tranquada {\it et al.}~\cite{Tra_nature,Tra_prb,Tra_prl_97, 
Ichikawa_00} observed elastic magnetic peaks at tetragonal positions
$(1/2\pm\epsilon, 1/2\pm\epsilon, 0)$ that are almost identical to those
of the inelastic IC peaks found in superconducting LSCO.  They explained
this static feature in terms of a two dimensional stripe model in which
hole-free antiferromagnetic (AF) regions are separated by one-dimensional
stripes of hole-rich regions.  Thus, their model places both
spin-density-wave (SDW) and charge-density-wave (CDW) order on the CuO$_2$
planes.  Since the charge stripes become the anti-phase boundaries of the
AF regions, the magnetic modulation period is twice that of the charge
density modulation.  In fact, additional satellite peaks were observed
by both neutron~\cite{Tra_prb,Ichikawa_00} and
X-ray~\cite{Zimmermann_98,Niemoller_99} diffraction techniques in LNSCO
around the nuclear Bragg peaks at
$(2\pm 2\epsilon, 0, l)$ due to the charge density modulation,
consistent with the stripe model as well as with more general coupled SDW
and CDW models.
Following these experiments, the same type of SDW order has been observed
by neutron scattering in LSCO samples with
$x=0.12$~\cite{Suzuki_98,Kimura_99} and in oxygen-doped stage-4
La$_2$CuO$_{4+\delta}$ (LCO($\delta$))~\cite{YSLee_99}.
Surprisingly, no charge order peaks have been detected yet in these
materials.  In all of the above cases, the between-plane correlation
length of the SDW order is of order or less than one lattice constant.

Since the stripe structure contains magnetic order, the effect of an
external magnetic field on the stripe should provide important
information about its nature.  To date, a few neutron scattering
experiments have been carried out under magnetic field to investigate
the effects on the SDW order in LSCO with $x=0.12$\cite{Katano_00}
and $x=0.10$,\cite{Lake_02} and in stage-4
LCO($\delta$).\cite{Khaykovich_02,Khaykovich_02b}  
All of these measurements show
qualitatively consistent behavior, with the SDW peaks being substantially
enhanced by applying a field perpendicular to the CuO$_2$ planes.
Possible explanations of the enhancement of the SDW peaks have involved
suppression of spin fluctuations or competing superconducting and AF
order~\cite{Arovas_97,Demler_01,Zhang_02,Kivelson_02,Zhu_02,Chen_02}
whose physical origin is the suppressed superconductivity
together with the enhanced AF order in the vortex cores. However, the
effects of a magnetic field on the stripe order itself remain unclarified.


In the present experiment, we have studied the effect of an applied
magnetic field on the stripe order in superconducting
La$_{1.45}$Nd$_{0.4}$Sr$_{0.15}$CuO$_{4}$.  The stripe order at this
particular composition has previously been characterized by
neutron\cite{Tra_prl_97} and X-ray\cite{Niemoller_99} diffraction and by
the zero-field muon-spin-relaxation ($\mu$SR) technique,\cite{Nachumi_98}
and the superconductivity has been characterized by high-field
magnetization measurements\cite{Ostensen_97} and by transverse-field
$\mu$SR.\cite{Nachumi_98}  We find that while an applied magnetic field of
$<1$~T is sufficient to suppress the parasitic ordering of Nd$^{3+}$
moments at the SDW wave vector, it has essentially no impact on the
stripe order associated with the doped holes and copper spins.

The rest of the paper is organized as follows.  After describing the
experimental details in the following section, the neutron and X-ray
scattering results are presented in Sec.~III.  These results are
discussed in Sec.~IV.  There we first explain the response of the Nd
moments in the magnetic field.  Then we consider the lack of effect on
the charge and Cu-spin stripes in the present sample, together with the
field-induced response in LSCO and LCO($\delta$).  Collectively, these
results can be understood in terms of the ideas 1) that there is little
coupling of a uniform magnetic field to the locally antiferromagnetic
correlations of the stripe phase and 2) that the suppression of
superconductivity in magnetic vortex cores results in the nucleation of
patches of stripe order.

\begin{figure}
\centerline{\epsfxsize=2.8in\epsfbox{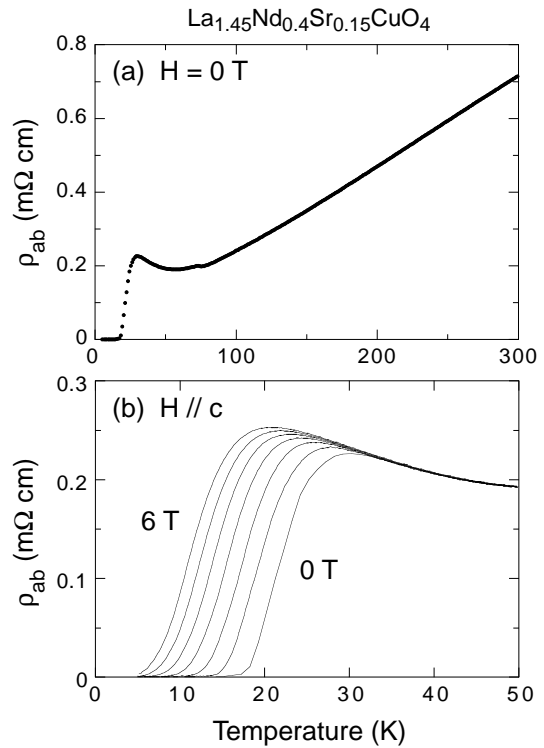}}
\caption{(a) Temperature dependence of the in-plane resistivity.
(b) Effect of a $c$-axis magnetic field on the in-plane resistivity.  
Data are shown for 0, 1, 2, 3, 4, 5, and 6 T.}
\end{figure}

\section{Experimental details}

The single crystal of LNSCO ($x=0.15$ and $y=0.4$) used in this study
is the same one used in Ref.~\onlinecite{Tra_prl_97}.  
The sample was grown using the
travelling-solvent floating-zone method and is 5 mm in diameter and
20 mm in length.  The sample exhibits a structural transition from a
low-temperature orthorhombic structure to a LTT structure at $\sim 80$~K;
the lattice
constants at 5~K are $a=b=3.80$~\AA~ and $c=13.1$~\AA, corresponding
to reciprocal lattice units of $a^*=b^*=1.65$~\AA$^{-1}$ and
$c^*=0.48$~\AA$^{-1}$.

The superconducting transition has been characterized on pieces of
crystal grown in the same fashion as the neutron sample.  From previously
reported measurements of the magnetic susceptibility\cite{Tra_prl_97} and
the thermodynamic critical field,\cite{Ostensen_97} the transition
temperature is approximately 10~K.  To characterize the effect of a
magnetic field applied along the $c$ axis on the transition, the
resistivity was measured, as shown in Fig.~1.  As is frequently observed,
the zero-field resistivity measurement indicates a higher transition
temperature than does the susceptibility.
We note the relatively large difference between $T_c$ values of the 
susceptibility and resistivity measurements in this system compared to 
the LSCO and LCO(d) systems.  Since the superconductivity in LNSCO is 
strongly suppressed by the CuO$_6$ octahedral tilt of the LTT structure 
which pins the stripes, the local $T_c$ is very sensitive to the local 
fluctuation in Sr and/or Nd concentration that cause the local 
fluctuation of the octahedral tilts.  
Bulk resistivity shows higher $T_c$ if there are small patches with 
higher $T_c$ that percolate through the sample.  On the other hand, 
if they have a small volume fraction, the magnetization will not be 
affected.  Thus, the relatively large difference of $T_c$ values may be a 
characteristic feature in LNSCO.  
However, this effect has no impact on the conclusions that we will 
reach based on the neutron and X-ray measurements.  

The neutron scattering experiments were performed using the triple axis
spectrometer SPINS installed on the cold neutron guide NG5 located at
the NIST Center for Neutron Research.  Highly-oriented pyrolytic graphite
crystals were used as monochromator and analyzer.  An incident neutron
energy of 5~meV with a horizontal collimation sequence
$32'$-$80'$-S-$80'$-open  (S : sample) was utilized.  Higher order
neutrons were removed from the  beam by a cold Be-filter located after the
sample.  The crystal was fixed  to an Al holder by Gd cement and Al wire,
and mounted in a cryostat  equipped with a superconducting magnet.  The
$a$ and $b$ crystallographic  axes were oriented in the horizontal plane
to allow access to $(h, k, 0)$  type reflections.  With this
configuration, the magnetic field was aligned  perpendicular to the
CuO$_2$ planes.  During the experiments, we verified  that the nuclear
Bragg intensities did not change with field, thereby  confirming that the
sample was properly mounted, that is, the sample  position was
field-independent.

The X-ray scattering experiments were carried out at the BW5 beam
line at HASYLAB in Hamburg, Germany.  The incident photon energy of
100 keV was selected by a Si$_{1-x}$Ge$_x$ gradient crystal monochromator
and analyzed by the same type of crystal.  The sample was mounted in
a superconducting magnet with the $c$-axis oriented perpendicular to the
scattering plane and the field aligned perpendicular to the CuO$_2$
planes. The momentum resolution (FWHM) measured at the $(2, 0, 0)$
Bragg position was 0.015~\AA$^{-1}$ along [100] and
0.0014~\AA$^{-1}$ along [010].

\begin{figure}
\centerline{\epsfxsize=2.8in\epsfbox{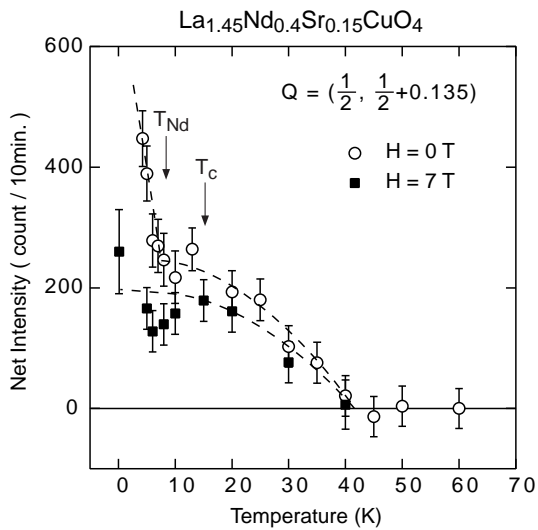}}
\caption{Temperature dependence of the net IC peak intensity at
$Q = (1/2, 1/2+0.135)$ in zero field (circles) and 7~T (squares).
Dashed lines are guides to the eye.   The data in a magnetic
field were measured on warming after field-cooling from 60~K.
In zero field, the intensity first appears below $\sim 40$~K and
grows rapidly below $T_{\rm Nd} \sim 8$~K due to the Nd$^{3+}$ ordering.
The rapid increase below $T_{\rm Nd}$ is suppressed at 7~T, but otherwise
the intensity appears to be constant within the errors.  This implies
that the magnetic field destroys the Nd$^{3+}$ ordering.  Above $T_{c}$
(or $T_{\rm Nd}$), there is at best a small diminution in intensity with
magnetic field.}
\end{figure}

\section{Neutron and X-ray cross-sections}

\subsection{SDW order}

In zero magnetic field, SDW IC peaks are observed at
$(1/2 \pm \epsilon, 1/2 \pm \epsilon, 0)$, where $\epsilon = 0.134$.
The temperature dependence of the SDW peak intensity is plotted in
Fig.~2 using open circles.  The peaks first appear at 40~K, which agrees
with the results of Ref.~\onlinecite{Tra_prl_97}.  
The peak intensity increases rapidly with
deceasing temperature below 8~K; this is due to the ordering of the
Nd$^{3+}$ spins.~\cite{Tra_prb}  Hereafter, we refer to this Nd ordering
temperature as $T_{\rm Nd} = 8$~K.  On the other hand, the temperature
dependence under magnetic field below $T_{\rm Nd}$ is significantly
different.  The squares in Fig.~2 represent the peak intensities
measured under a 7~T magnetic field.  Although there seems to be a small
reduction of intensity, the temperature dependence above $T_{\rm Nd}$ is
quite similar to that in zero field.  Importantly, however, there is no
longer a rapid increase in intensity below $T_{\rm Nd}$ at 7~T.
%
\begin{figure}
\centerline{\epsfxsize=2.8in\epsfbox{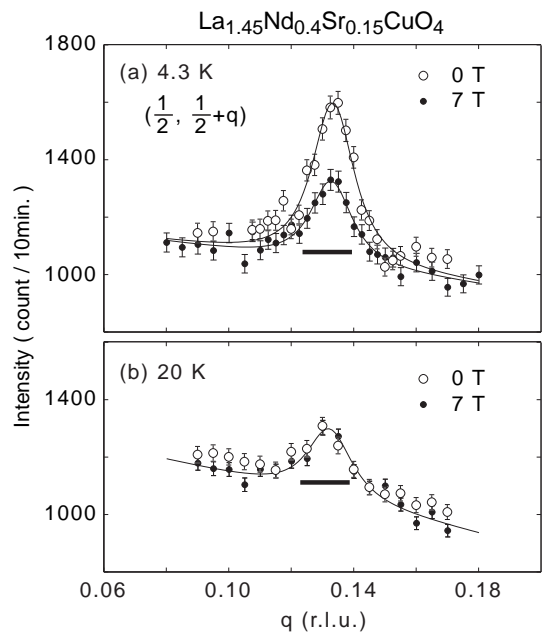}}
\caption{Lineshape of the $(1/2, 1/2+\epsilon, 0)$ IC peak at (a) 4.3~K
and (b) 20~K in zero field (open circles) and 7~T (closed circles).
(Only the open circle is shown when the symbols overlap.)  The data in
the 7~T magnetic field were taken after field-cooling from 60K.
Horizontal bars indicate the instrumental resolution.  Solid lines are
the results of fits to a two-dimensional Lorentzian function convoluted
with the instrumental resolution.  A clear reduction of the IC peak
intensity is observed at $4.3$~K (below $T_{\rm Nd}$), while no
significant  change occurs at 20~K (above $T_{\rm Nd}$).}
\end{figure}
%
These features are more clearly shown in Fig.~3, which shows peak profiles
measured along $(1/2, 1/2+q, 0)$ at 4.3~K and 20~K.  At 4.3~K (below
$T_{\rm Nd}$), the peak intensity at 7~T is reduced to half of that in
zero  field, while there is no significant change with field in the peak
profile  at 20~K (above $T_{\rm Nd}$).  For all profiles, the peak widths
are slightly  larger than the instrumental resolution width which is
represented by the  thick horizontal bars.  The solid lines in Fig.~3 are
the results of fits  to a resolution-convoluted two-dimensional (2D)
Lorentzian function of $q$. These fits show that the correlation length
of the SDW order is $\sim  200$~\AA~ for all profiles, that is, only the
intensity changes with  temperature and magnetic field.

\begin{figure}
\centerline{\epsfxsize=2.8in\epsfbox{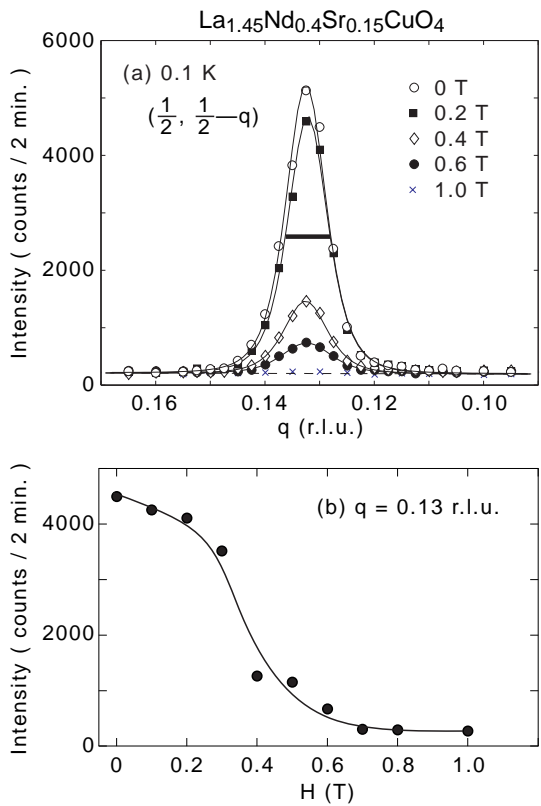}}
\caption{(a) IC peak profiles at 0.1~K in different magnetic fields.
The horizontal bar indicates the instrumental resolution.  The solid lines
are the results of fits to a two-dimensional Lorentzian-squared function
convoluted with the instrumental resolution.  (b) Field dependence of the
IC peak intensity at 0.1~K.  The solid line is a guide to the eye.  Most
of the intensity is suppressed by a magnetic field of 0.7~T.  However,
even at 7~T a weak intensity comparable to that observed above 8~K remains
as shown in Fig.~2.}
\end{figure}

The intensity of the SDW peak in zero field increases by more than an
order of magnitude below $T_{\rm Nd}$ as indicated in Fig.~2.
We find that the field-induced suppression of the SDW peaks is especially
significant at the lowest temperatures.
Figure~4(a) shows the field dependence of the SDW peak measured at
$(1/2, 1/2-\epsilon, 0)$ at 0.1~K.  In contrast to the factor of 2
reduction caused by 7~T at 4.3~K, the peak intensity is almost
completely suppressed by a field of less than 1~T.  The solid lines
are the  results of fits to a resolution-convoluted 2D Lorentzian squared
function of $q$.  The peak width is almost resolution-limited and
does not change with field.  The field dependence of the peak
intensity is shown in Fig.~4(b).  The intensity decreases rapidly
with increasing field and almost reaches the background at $H=0.7$~T.
Although the intensity appears to be completely suppressed at $H \geq
0.7$~T, there is still a small remaining signal that is comparable to
that observed just above $T_{\rm Nd}$.  This intensity is shown in Fig.~2
by the solid square at the lowest temperature.

\subsection{CDW order}

The temperature dependence shown in Fig.~2 naturally
suggests that the applied field primarily suppresses the Nd spin contribution
to the SDW peaks.  In particular, the temperature dependence at 7~T above
$T_{\rm Nd}$ is very close to that in zero field, and the drastic increase
of the  SDW peak intensity below $T_{\rm Nd}$ disappears at 7~T.  The next
question  is then how does the field affect the CDW peaks?  To study
this, we  performed X-ray scattering experiments in an applied magnetic
field.
Figure~5 shows the temperature dependence of the CDW peak intensity
measured at $(1.74, 0, -0.5)$.  Circles and diamonds
correspond to data taken in zero field and 4~T, respectively.
The choice of $L=-0.5$ was made because the structure factor has a
maximum at that position.~\cite{Zimmermann_98}
The onset temperature is $\sim 60$~K which
is consistent with the previous measurement for $x=0.15$ in 
Ref.~\onlinecite{Zimmermann_98},
and is same as that reported for the $x=0.12$
sample.~\cite{Tra_prb,Zimmermann_98}  As shown in Fig.~5, the temperature
dependences with and without field are completely identical.  The inset
shows the field dependence of the intensity at 1.9~K and 4~K.  We find,
therefore, that while the SDW peak is strongly suppressed by application
of a magnetic field below $T_{\rm Nd}$, the CDW peak intensity is
independent  of field.

\begin{figure}
\centerline{\epsfxsize=2.8in\epsfbox{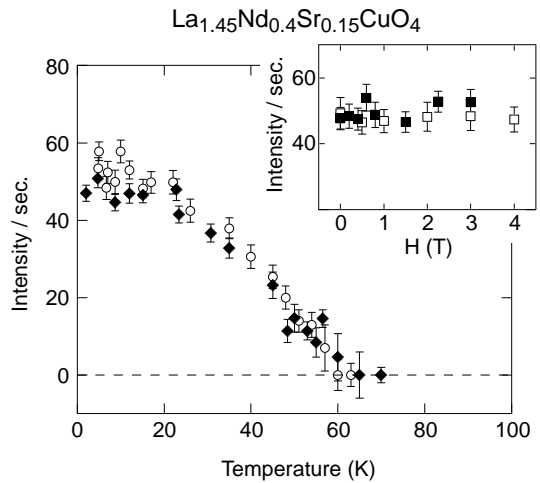}}
\caption{Temperature dependence of the CDW peak intensity measured
at $(1.74, 0, -0.5)$ in zero field (circles) and in $4$~T
(diamonds).  The inset shows the field dependence of the peak
intensity at 1.9~K (open squares) and 4~K (closed squares).}
\end{figure}

\section{Discussion}

\subsection{Nd response}

 From these results,
it appears that the magnetic field perpendicular to the
CuO$_2$ planes inhibits the Nd$^{3+}$ spin ordering, but has at most a
weak effect on the stripe structure itself.  This picture confirms that
the Nd spins simply follow the stripe order of the Cu spins and not the
other way around.  It also suggests that the correlation between Cu and
Nd spins is weak as observed previously in the related material
Nd$_{2}$CuO$_{4}$.~\cite{Matsuda_90}  Below we discuss the Nd response in
more detail.

The Nd spin contribution to the SDW peak intensity is dominant at the
lowest temperatures\cite{Tra_prb}; therefore, the field dependence in
Fig.~4(b) should relate to the magnetic response of the Nd ions.  The
magnetic fluctuations of the Nd ions in La$_{2-x-y}$Nd$_y$Sr$_x$CuO$_4$
have been studied with neutron scattering by Roepke {\it et
al.}~\cite{Roepke_99} For a sample with $y=0.3$ and $x=0$ measured at low
temperature, they resolved an excitation at 0.25 meV which they
attributed to a splitting of the Kramers-doublet ground state of
Nd$^{3+}$ by an exchange interaction with ordered Cu moments.  In a
sample with $y=0.6$ and
$x=0.15$, the magnetic fluctuations appeared as quasi-elastic scattering
with a half-width of $\Gamma/2 = 0.1$~meV.  If we take this energy width
to represent the effective exchange interaction appropriate for our
sample ($y=0.4$), then the external field that is required to give an
equal Zeeman energy is $H_{0}=\Gamma/2m_{\rm Nd}$,
where $m_{\rm Nd}$ is the magnetic moment of a Nd ion.  From the
magnetization measurements of Ostenson {\it et al.}\cite{Ostensen_97} on
a crystal identical to ours,
$m_{\rm Nd}$ is 3.2~$\mu_{\rm B}$, which finally  gives
$H_{0} = 0.54$~T.  This estimate is in good agreement with the
field at which the peak intensity drops, as shown in Fig.~4.   We conclude
that a modest uniform magnetic field is sufficient to align the Nd moments
uniformly, thus removing their contribution from the SDW superlattice
peaks.

The dominant part of the Nd contribution to the SDW peaks appears below
$T_{\rm Nd}$; however, there may also be a small contribution from Nd
moments at higher temperatures.\cite{Tra_prb}  The small decrease of the
SDW peak intensity caused by the 7~T field for $T>T_{\rm Nd}$ (see Fig.~2)
is likely due to the elimination of the Nd component.  Note that the CDW
intensity shows no significant change for applied fields up to 4~T.

\subsection{Magnetic field, superconductivity, and stripes}

In a recent $\mu$SR study\cite{Savici_02} on LSCO with $x=0.12$ and
stage-4 LCO($\delta$), it was found that local magnetic (SDW) order
occurred in only a fraction of the volume, 20\%\ and 40\%, respectively.
Within that volume fraction, the average local hyperfine field is the
same as in a uniformly stripe-ordered sample,
La$_{1.48}$Nd$_{0.4}$Sr$_{0.12}$CuO$_4$.~\cite{Nachumi_98}  
Based on neutron diffraction
studies,\cite{Lake_02,Kimura_99} the volume fraction exhibiting SDW
order in LSCO with $x=0.10$ should be much smaller.
LSCO\cite{Katano_00,Lake_02} and LCO($\delta$)\cite{Khaykovich_02} show
a clear enhancement of the SDW peaks in the presence of a field applied
perpendicular to the CuO$_2$ planes.  This enhancement could result from
a coupling of the magnetic field to the SDW order parameter, or from
growth of the SDW volume fraction by suppression of the superconductivity.

A direct coupling to the SDW order has recently been observed in LSCO
with $x=0.024$ by Matsuda {\it et al.}\cite{Matsuda_02}  This sample is
insulating at low temperatures and exhibits diagonal IC SDW order.  A
small reduction of the SDW peak intensity was found in an applied
magnetic field.  The effect was explained in terms of the reorientation
of spins in half of the CuO$_2$ layers in order to align the canted spin
components that result from the Dzyaloshinskii-Moriya exchange
interaction.  In the case of the parallel (vertical) stripes present in
the superconducting phase, we would expect the orientation of the canted
spin components to alternate in neighboring magnetic domains so that
there would be no net coupling to a uniform field.  The long correlation
length observed in LNSCO vitiates the mechanism of Matsuda {\it et
al.}\cite{Matsuda_02}  Furthermore, such a mechanism would not explain
the field-induced {\it enhancement} in LSCO and LCO($\delta$).

The present results indicate that there is no significant direct coupling
between a uniform magnetic field and the stripe order.  This is
consistent with expectations, given that the magnetic order is locally
antiferromagnetic.  It follows then that the field-induced growth of SDW
order in LSCO and LCO($\delta$) must be due to suppression of the
superconductivity.  The superconductivity may coexist with SDW order in
regions of these samples, but the $\mu$SR results indicate that there
must be a significant volume fraction where there is superconductivity
without SDW order.  It is presumably in these latter regions that new SDW
order is generated by the applied field.\cite{Khaykovich_02b}  It is then
understandable that the largest SDW growth with field occurs in LSCO with
$x=0.10$,~\cite{Lake_02} 
where the zero-field SDW volume fraction is quite small.
It is also reasonable that there is no significant enhancement in the 
present crystal which is reported to be unifirmly SDW ordered by 
$\mu$SR.~\cite{Nachumi_98}

The applied field penetrates the sample in quantized vortices, with the
superconducting order parameter going to zero within each vortex core.
The possibility that N\'eel order might appear in the vortex cores was
considered by Arovas {\it et al.}\cite{Arovas_97}  A more relevant model,
in which SDW and superconducting order can coexist, was analyzed by
Demler {\it et al.},\cite{Demler_01,Zhang_02} with some refinements
proposed later by Kivelson {\it et al.}\cite{Kivelson_02} (see also
Refs.~\onlinecite{Zhu_02,Chen_02}).  In this model, SDW order can be
induced in a region extending beyond the vortex core, similar to the
``halo'' effect observed by scanning tunneling
microscopy\cite{Hoffman_02} on Bi$_2$Sr$_2$CaCu$_2$O$_{8+\delta}$.  The
long magnetic correlation lengths observed at high field in
LSCO\cite{Katano_00,Lake_02} and LCO($\delta$)\cite{Khaykovich_02}
indicate that the halo radius may be
$>100$~\AA.  It seems likely that weak Ising anisotropy, which is known
to be present at low doping,\cite{Matsuda_02} is important for
establishing static order in domains of finite extent.  Based on the
present results, we would expect the SDW order to saturate when the
vortex spacing is comparable to the halo diameter.

Finally, continuing with the same argument, the lack of a significant
change in the stripe order in our LNSCO sample provides further
confirmation that the stripe order is uniform in the sample consistent 
with the $\mu$SR measurement.~\cite{Nachumi_98}  If it were
not, then there should be regions with superconductivity and no SDW
order, and we would expect to see an increase in the SDW order as
vortices are induced in those regions.  This result also confirms that
the bulk superconductivity\cite{Ostensen_97,Nachumi_98} must coexist
with stripe order.  
This is also supported by the fact that there is no significant anomaly 
in SDW and CDW orders at $T_c$ in this material.

\begin{acknowledgments}

We thank P. M. Gehring, B. Khaykovich, S. Park, G. Shirane, J. R. Schneider,
M. Matsuda, and C. Broholm for invaluable discussions.  We also acknowledge
Thomas Br\"uckel for allowing us to use his magnet during the X-ray
scattering experiment.  The work at the University of Toronto is part of the
Canadian Institute for Advanced Research and supported by the Natural Science
and Engineering Research Council of Canada.  Research at Brookhaven National
Laboratory was carried out under Contract No.\ DE-AC02-98CH10886, Division of
Materials Sciences, U.\ S. Department of Energy, while research at the
University of Tokyo was financially supported by a Grant-in-Aid for
Scientific Research from the Ministry of Education, Science, Sports and
Culture of Japan.  Finally the work at SPINS in National Institute of Standards
and Technology is based upon activities supported by the National Science
Foundation under Agreement No. DMR-9986442.

\end{acknowledgments}

\end{document}